\documentclass[12pt,a4paper]{article}
\usepackage{amsmath,amsfonts,amssymb}
\usepackage[dvips]{graphicx}
\begin{document}
\title{Can the Stephani model be an alternative to FRW accelerating models?}
\bigskip
\author{W\l odzimierz God{\l}owski${^1}$ \&
Jerzy Stelmach${^2}$ \&
Marek Szyd{\l}owski ${^1}$}
\maketitle
1. Astronomical Observatory of the Jagiellonian University, 30-244
Krakow, Orla 171, Poland

2. Institute of Physics, University of Szczecin, Wielkopolska 15,
70-451 Szczecin, Poland  e-mail: jerzy.stelmach@univ.szczecin.pl
\bigskip
\section*{Abstract}

\medskip

A class of Stephani cosmological models as a prototype of non-homogeneous
universe is considered. The non-homogeneity can lead to accelerated evolution
which is now observed from the SNIa data. Three samples of type Ia supernovae
obtained by Perlmutter {\it at al.}, Tonry {\it et al.} and Knop {\it et al.} are
taken into account.
Different statistical methods (best fits as well as maximum likelihood method)
to obtain estimates of the model parameters are used. Stephani model is considered
as an alternative to the concordance of $\Lambda$CDM model in the explanation of
the present acceleration of the universe. The model explains the acceleration
of the universe at the same level of accuracy as the $\Lambda$CDM  model ($\chi^2$
statistics are comparable). From the best fit analysis it follows that the Stephani
model is characterized by higher value of density parameter $\Omega_{m 0}$ than the
$\Lambda$CDM model. It is also shown that the obtained results are consistent with
location of CMB peaks.

\section{Introduction}

In the paper of Stelmach and Jakacka \cite{jak01} it was suggested that the
effect of acceleration of the universe can be driven by non-homogeneities
in the Stephani model \cite{kra85,kra97}. We perform the statistical
verification of this hipothesis using current available astronomical
data. We use three different samples of supernovae where Perlmutter sample
\cite{Perlmutter99} is treated as the fiducial data set. For comparison both the
Tonry {\it et al.} \cite{Tonry03} (improved by Barris {\it et al.} \cite{Barris03})
and Knop {\it et al.} \cite{Knop03} are used.

To find which model fits to the data in the best way we use the $\chi^2$ statistics.
The parameters of the models are estimated using the maximum likelihood method.
One-dimensional probability distribution function (pdf) over model parameters is
presented to deepen statistical insight.

Recently found accelerated expansion of our Universe \cite{sup98} is explained
in the literature by the presence of the $\Lambda$-term or the so-called
quintessence matter satisfying the equation of state with negative pressure
\cite{stein98}. Observation of type Ia
supernovae (SN Ia) revealed that this form of energy called dark energy
could dominate the present evolution of the universe. Therefore it
is important to consider different candidates for dark energy. In the paper of
Stelmach and Jakacka it was shown that non-homogeneity in spherically symmetric
Stephani model can be treated as some kind of dark energy due to which the
universe accelerates (we call this model ``S-J model'').
It is a sort of fictitious fluid for which density parameter $\Omega_{\rm non}$
can be defined. Then dynamics can be formally reduced to the FRW flat model
with additional fluid satisfying the  equation of state $p=(-2/3)\rho$
corresponding to the equation of state for topological defects in a form of domain walls
\cite{IAU87, AJII,stelbd93}.

The best fitting procedure applied to supernovae data as well as confidence
level for the redshift-magnitude relation shows that $\Omega_{\rm non}$ should
dynamically dominate the present evolution of the Universe if we want to explain
observational data without cosmological term (see also \cite{Celerier00}).

\section{Stephani Model in the S-J version}

Stephani model in the S-J version can be described by the generalized Friedman
equation \cite{jak01}
\begin{equation}
\dot {R}^2+k=\frac{A^2} {R^{1+3\alpha}},
\end{equation}
where $k(t)=\beta R(t)$ is an effective curvature index which depends on time.
$\beta$ and $\alpha$ are constants ($p=\alpha\rho$). In the above equation
non-homogeneity is not explicitly present. This is due to the choice of the
time parameter.

Equation (1) is de facto a first integral of the Einstein equation which is a
second order equation for the scale factor. It is obtained by assuming that
the observer is placed at a symmetry center $p(r\approx 0, t)=
\alpha\rho(t)$. We consider the universe in the neighbourhood of this
center.

The relation
\begin{equation}
k(t)=\beta R(t),
\end{equation}
is a special form of a more generalized ansatz of a type
\begin{equation}
k(t)=\beta R^{\gamma}(t), \gamma= {\rm const.}
\end{equation}
However, $\gamma=1$ is a simplest case, and our goal is to show that such
exotic model can explain SN Ia data of Perlmutter even in this simple case.

For further purposes it is convenient to write down Eq. (1) in a
dimensionless form
\begin{equation}
{\dot{x}^2}=\sum\limits_{i=0}^1\Omega_{i,0}x^{-\left(3\alpha_i+1\right)},
\end{equation}
where $x\equiv R/R_0$ is the so-called radius of the universe in units
$R_0$. Here the differentiation is carried out over the dimensionless
time parameter $\tau \equiv |H_0|t$. The parameters $\Omega_{i 0}$ ($i=0,1$)
are defined
\begin{equation}
\Omega_{i 0}=\frac{\rho_{i 0}}{3{H_0}^2},
\end{equation}
where the subscript $0$ means that a quantity with this subscript corresponds
to the present epoch
\begin{equation}
\rho_{i 0}\equiv \rho_i(x=1).
\end{equation}
In our model $\alpha_0\equiv \alpha$ and $\alpha_1\equiv -2/3$
like for topological defects, hence
\begin{equation}
\rho_\alpha\equiv\frac{C_{\alpha}}{R^{3+3\alpha}}, ~~{\rm and}~~
\rho_{\rm non}\equiv\frac{(-3\beta)}{R},
\end{equation}
where $C_{\alpha}$ is some constant.

In the given parametrization our non-homogeneous Stephani model looks formally
as a flat Friedman model with two noninteracting components.

\section{Stephani model as a hamiltonian system}

Because of interpretation of the S-J model in terms of FRW model with additional
fictious fluid we can simply find hamiltonian formalism for dynamics of the
system. A first integral of the Friedman equation can always be used for reduction
of the motion of the system to a motion of a particle of unit
mass in one dimensional potential \cite{dab86}. If we define
\begin{equation}
\rho_{\rm eff}\equiv \sum\limits_i \rho_i(R)= f(R),
\end{equation}
then the potential is
\begin{equation}
V(R)\equiv -\frac{\rho_{\rm eff}} {6}R^2=\frac{\beta}{2}R-\frac{A^2} {2}
\frac{1} {R^{1+3\alpha}}
\end{equation}
and the Euler-Lagrange equation of motion reads
\begin{equation}
\ddot R=-\frac{\partial V}{\partial R}
\end{equation}
and has a Newton-like form.

The Hamiltonian in the considered case has the form:
\begin{equation}
{\cal H}=\frac{P_R^2}{2} +V(R),
\end{equation}
where we defined $P_R\equiv\dot {R}$

Trajectories of the system lie on a zero energy level ${\cal H} = 0$ and
a hamiltonian constraint must coincide with the form of the first integral
of the equation of motion (10).

The advantage of having the hamiltonian function given by Eq (11) is that
one is then able to make easily full classification of solutions in the
configuration space \cite{jak01}. Moreover dynamics can be presented
in two dimensional phase plane.

The corresponding equations of motion describe a simple dynamical system
consisting of a particle moving in one dimensional potential:
$$\dot{x}=y, \nonumber$$
\begin{equation}
\dot{y}=-\frac{\partial V} {\partial x},
\end{equation}
where $x=R$, $y=\dot{R}$ and $V(x)$ is given by (9),
and $y^2/2 +V(x)=0$ is its first integral.
Qualitative analysis of differential equation implies
a shift from finding and analysing individual solution to investigating
the space of all solutions. Certain properties (such as existence or absence of
horizons, existence of singularities) are believed to be realistic if they can
be attributed to a larger class of models within the space off all solutions
(phase space).

This approach offers a possibility of investigating the space of all possible
solutions for the considered problem. Of course the system (12) possesses the
first integral ${\cal H}=0$ which defines a family of algebraic curves (phase
curves) on which the trajectories of the system lie. The phase portrait of
the considered system for a) $\alpha=0$ (dust) is shown in Fig. 1 and for
b) $\alpha=1/3$ (radiation) in Fig. 2.

\begin{figure}[h]
\begin{center}
\includegraphics[angle=270, width=0.57\textwidth]{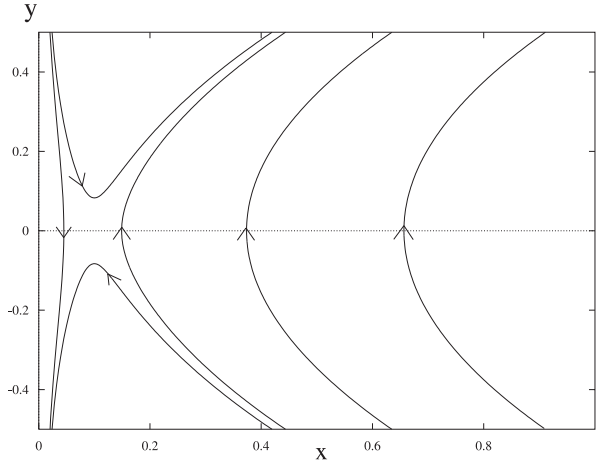}
\caption{Phase portrait of S-J model for dust.}
\label{fig:1}
\end{center}
\vspace{-0.6cm}
\end{figure}

\begin{figure}[h]
\begin{center}
\includegraphics[angle=270, width=0.57\textwidth]{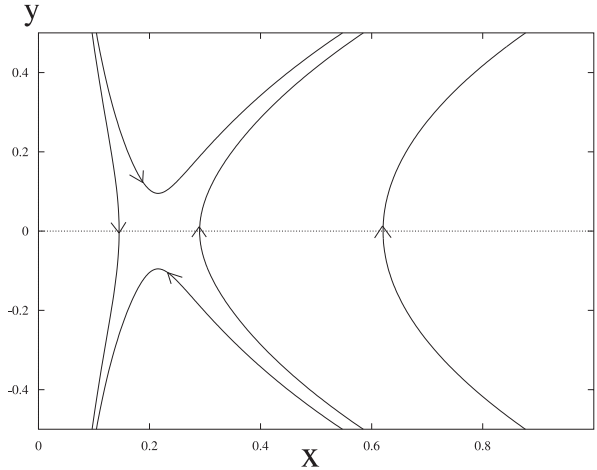}
\caption{Phase portrait of S-J model for radiation.}
\label{fig:2}
\end{center}
\vspace{-0.6cm}
\end{figure}

The system is described by the equations $\dot{x}=y$,
$\dot{y}=-{\partial V / \partial x}$ and the first integral is
${y^2/2}+V(x)=0$, where $V(x)=(1/2)(\beta x-{A^2/x^{1+3\alpha}})$ and we put
$\beta=-1$. The phase domain is $x > 0$. The acceleration region is situated
to the right of the saddle point, therefore for the Lema\^{i}tre-Eddington
(loitering) universes the acceleration begins in the middle of a quasi-static stage.

For comparison the FRW model with dust and cosmological constant is presented in Fig. 3.
We can observe how non-homogeneity term can mimic cosmological constant term.
In the particular cases we have:
a) $\dot{x}=y, \dot{y}=-(1/2)(\beta x +A^2/x^{2})$,
b) $\dot{x}=y, \dot{y}=-(1/2)(\beta +A^2/x^{3})$, where $A=0.1, \beta=-1,$
c) $\dot{x}=y, \dot{y}=-(\Lambda/3)x-\rho_0/6x^{2}$,
where $\Lambda=0.7, \rho_0=0.3$ with $V(x)= -(\Lambda/3)x^2-\rho_0/6x.$

\begin{figure}[h]
\begin{center}
\includegraphics[angle=270, width=0.57\textwidth]{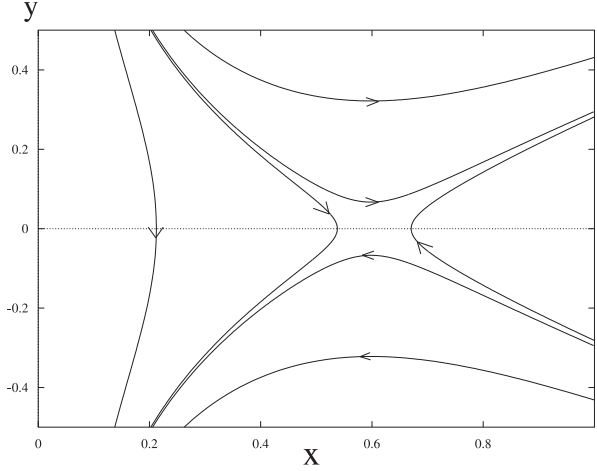}
\caption{Phase portrait of FRW model with dust and cosmological constant.}
\label{fig:3}
\end{center}
\vspace{-0.6cm}
\end{figure}

The phase portrait is organized by critical points ($x_0, y_0$) (i.e singular
solution for which $y_0=0$ and ${\partial V /\partial x}|_{(x_0,y_0)}=0$) and
joining them trajectories. Hence the phase portrait provides
a global qualitative picture for the dynamics. Two phase portraits are equivalent
if there exists an orientation preserving homeomorphism mapping integral curves
of the systems. In our case the critical point for $\beta +(1+3\alpha)<0$
is representing by static universe. Then
\begin{equation}
y_0=0, \qquad x_0=\left(-\frac{\beta} {(1+3\alpha)}A^2\right)^{-\frac{1}{(2+
3\alpha)}},
\end{equation}
where $\beta/(1+3\alpha)<0.$
While constructing the phase portrait it is important for the system to be
linearized near the critical points because the Hartman-Grobman theorem
says that the original system is equivalent to its linear part in the nearby
hyperbolic critical points (a critical point is hyperbolic (non-degenerate))
if there exists such $i$ that ${\rm Re}\lambda_i\ne 0$, where $\lambda_i$ are
eigenvalues of a linearization matrix.
A character of critical points is determined by characteristic
equation for the linearization matrix
\begin{equation}
A={0, \qquad 1 \choose - \frac{\partial^2 V}{\partial x^2},
\qquad 0}_{(x_0,y_0=0)}.
\end{equation}
Since in our case ${\rm Tr} A=0$ only saddle points
(${\partial^2 V/\partial x^2}|_{(x_0,0)}<0$)
or centres (${\partial^2 V/\partial x^2}|_{(x_0,0)}>0$) are admissible.

After some simple calculations one can check that a type of the critical point
depends on the diagram of the potential, namely:
\begin{equation}
{\rm det}A=\frac{\partial^2 V}{\partial x^2}|_{(x_0,0)}=-\frac{1}{2}A^2(1+3\alpha)
(2+3\alpha)R_0^{-3(1+\alpha)}.
\end{equation}
Therefore for $\alpha>1/3$ there exists only one critical point which is of
saddle type, i.e it is a structurally stable point. The full knowledge of the
dynamical system comprises also its behaviour at infinity. To achieve this
one usually transforms the phase space into a Poincar\'e sphere. Then infinitely
distant points of the phase plane are mapped onto the sphere's equator. The phase
trajectories are mapped into corresponding curves on $S^2$. The character of
critical points is preserved and new critical points representing asymptotic
states of the system can appear at the equator. Then an orthogonal projection
of any hemisphere onto a tangent plane compactifies phase portrait.

\section{Horizon and flatness problem}

Due to existence of the first integral in such a system we can discuss
some interesting properties of the evolutional path. There is a theorem
\cite{skd02} about nonexistence of particle horizon in such a model, namely
if $R \to 0$ and $\dot R < C$, where $C$ is a constant, then there is no
particle horizon in the past. In terms of potentials, if $V(R) \to {\rm
constant}$ for $R\to 0$ then there exists no particle horizon. It follows from
the implication
\begin{equation}
\dot R \le C \Rightarrow V \ge -{C^2\over 2}.
\end{equation}
We can see from Fig. 1 and Fig. 2 in the S-J paper that all models with
exception of case a) do not possess particle horizons. This is equivalent to
nonexistence of the horizon problem in the corresponding models. However, it
does not necessarily mean that if the condition $\dot R<C$ for $R\to 0$ is not
satisfied then the horizon problem appears. In that case explicit evaluation
of the integrals is required. We shall not deal with the problem in the present
paper.

As regards a flatness problem it does not appear in a cosmological model if
for later times the curvature term does not dominate the matter term. In our
model for the the problem not to appear we need the following condition to be
satisfied
\begin{equation}
|\beta R(t)|=|k(t)| \le {A^2 \over |R^{1+3\alpha}(t)|}.
\end{equation}
Hence
$$\alpha \le -{2\over 3}.$$
Of course this is not the case considered either in the S-J or in the present
paper, hence the flatness problem must be solved in a similar way as in the
standard cosmology (e.g. using inflationary scenario).

\section{Acceleration}

The case $\beta<0$ is particularly interesting because the corresponding
cosmological model without matter ($A=0$) evolves with constant acceleration.
With matter it accelerates always for later times independently of $\alpha$.
It follows from the relation
\begin{equation}
\ddot R=-{\partial V\over \partial R}= {1\over 2}\beta-{A^2\over 2}
(1+3\alpha)R^{-3\alpha-2}.
\end{equation}
Let us now differentiate both sides of the Eq. (4) with respect to $\tau$
\begin{equation}
\ddot x=-{1\over 2}\sum\limits_{i=0}^1(1+3\alpha_i)\Omega_{i 0}
x^{-\left(3\alpha_i+2\right)}.
\end{equation}
If we want the model to accelerate at the present epoch we must put
$\ddot x>0$ at $x=1$. Hence we get the expected result
\begin{equation}
\Omega_{{\rm non} 0}>\Omega_{\alpha 0},
\end{equation}
since this is the nonhomogeneity, which drives the acceleration.

\section{Magnitude-redshift relation}

It is well known that luminosity of observed objects, depends sensitively on
the spatial geometry (curvature) and dynamics of the Universe. Therefore
a cosmic distance measure, e.g. luminosity distance, depends on
the present densities of different components filling up the universe and
their equations of state. For this reason, the magnitude-redshift relation
is proposed as a potential test for cosmological models and play important
role in determining cosmological parameters \cite{dabh98}.

Let us consider an observer located at $r=0$ and at the moment $t=t_0$
receiving light emitted at $t$ from the source of absolute luminosity $L$
located at the radial distance $r$. Cosmological redshift $z$ of the source
is related to $t$ and $t_0$ by the formula
\begin{equation}
{1+z\over V(r,t)}={R(t_0)\over R(t)},
\end{equation}
where the function $V(r,t)$ is
\begin{equation}
V(r,t)=1+{1\over 4}\beta R(t)r^2.
\end{equation}
If the apparent luminosity of the source measured by the observer is $l$,
then the luminosity distance $d_L$ of the source, defined
\begin{equation}
l={L\over 4\pi d_L^2},
\end{equation}
is given by:
\begin{equation}
d_L={(1+z)R_0r\over V(r,t)}.
\end{equation}
For historical reasons, the observed and absolute luminosities are
defined respectively in terms of $K$-corrected observed and absolute
magnitudes $m$ and $M$ ($l=10^{-2m/5}\times 2.52 \times 10^{-5}
{\rm erg\,cm}^{-2}\,{\rm s}^{-2}$, $L=10^{-2M/5}\times 3.02 \times 10^{35}
{\rm erg\,s}^{-2}$) \cite{wein72}. When written in terms of $m$ and $M$,
Eq.(23) yields
\begin{equation}
m(z,{\cal M},\Omega_{\alpha 0})={\cal M} + 5\log_{10}
[{\cal D}_L(z,\Omega_{\alpha 0})],
\end{equation}
where
\begin{equation}
{\cal M}=M-5\log_{10}H_0+25
\end{equation}
and
\begin{equation}
{\cal D}_L((z,\Omega_{\alpha 0})\equiv H_0 d_L((z,\Omega_{\alpha 0},H_0)
\end{equation}
is a dimensionless luminosity distance in Mpc.

The radial coordinate $r$ in the expression (22) for $d_L$ can be found by
evaluation of the integral
\begin{equation}
r={1\over R_0H_0}\int\limits_x^1{dy\over \sqrt{\Omega_{\alpha 0}y^{1-3\alpha}+
(1-\Omega_{\alpha 0})y^3}},
\end{equation}
where $x\equiv R/R_0$. Since in order to compare with observations the
apparent luminosity $m$ must be a function of the redshift $z$, the $x$
parameter should be calculated from the relation
\begin{equation}
z(x)={1\over x}-1+{\Omega_{\alpha 0}-1\over 4}\left[{\int\limits_x^1{dy\over
\sqrt{\Omega_{\alpha 0}y^{1-3\alpha}+(1-\Omega_{\alpha 0})y^3}}}\right]^2.
\end{equation}
which follows from (21). Of course in a general case to find $x$ analytically
is not an easy task. Formally we have to find an inverse function $x(z)$.
The problem of finding $d_L(z)$ must be solved numerically and the simplest
way to do that is to regard the function $d_L(z)$ as given in a parametric
representation
\begin{equation}
d_L(x)={1\over xH_0}\sqrt{4\left[z(x)+1-1/x\right]\over \Omega_{\alpha 0}-1},
\end{equation}
and $z(x)$ is given by (29).

Similarly as in the paper of Jakacka and Stelmach we consider physically
the simplest case, i.e. we assume that in the neighbourhood of the symmetry
center the matter filling up the universe is a dust, hence $\alpha=0$. We define
$\Omega_{m 0}\equiv\Omega_{\alpha 0}|_{\alpha=0}$. In
this case the expression (29) takes a form
\begin{equation}
z(x)={1\over x}-1+{\Omega_{m 0}-1\over 4}\left[{\int\limits_x^1{dy\over
\sqrt{\Omega_{m 0}y+(1-\Omega_{m 0})y^3}} }\right]^2
\end{equation}
($\Omega_{{\rm non} 0}=1-\Omega_{m 0}$),
which can be evaluated using Weierstrass elliptic function $\cal P$
\cite{dab86}. The invariants $g_2$ and $g_3$ determining $\cal P$ function
are
$$g_2={1\over 4}\Omega_{m 0}(\Omega_{m 0}-1),\quad g_3=0.$$

Accuracy of the fit is characterized by the parameter
\begin{equation}
\chi^{2}=\sum_{i} \frac{(m_{0,i}^{\rm obs}-m_{0,i}^{\rm theor})^2}{\sigma_{m,i}^{2}
+ \sigma_{z,i}^{2}},
\end{equation}
where $m_{0,i}^{\rm obs}$ is the measured value, $m_{0,i}^{\rm theor}$
is the value calculated in the model described above, $\sigma_{m,i}^{2}$
is a measurement error of $m$, while $\sigma_{z,i}^{2}$ is a measurement error
of $z$ following from the dispersion in peculiar velocities of galaxies.

We assume that supernovae measurements come with uncorrelated Gaussian
errors and in this case the likelihood functions $\mathcal{L}$ can be
determined from chi-squared statistics
$\mathcal{L} \propto \exp{(-\chi^{2}/2)}$ \cite{Perlmutter99,Riess98}.

\section{Results of the statistical analysis with the Perlmutter sample.}

In the paper of Stelmach and Jakacka it was shown that spherically symmetric
Stephani cosmological model satisfying natural assumptions concerning local
equation of state for the matter may explain supernovae data of Perlmutter in
the context of the magnitude-redshift relation. In the present paper we
discuss this problem in more detail and we find numerical values of these
parameters such that observational data are best fitted to our theoretical
$m-z$ curve. We take the whole sample of Perlmutter (all 60 galaxies) into
account not rejecting any. Separately we analyse the sample of 54
supernovae (Perlmutter sample C), where 4 outliers and 2 reddened supernovae
were excluded, finding no significant differences.
In Fig.~4 we present plots of magnitude-redshift relations for the Stephani
model and the Perlmutter sample A.

\begin{figure}[h]
\begin{center}
\includegraphics[width=0.75\textwidth]{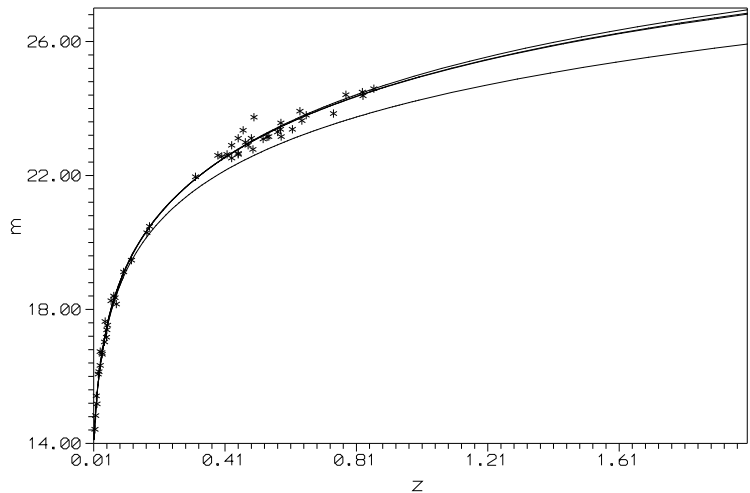}
\vspace{-0.1cm}
\caption{The Redshift-magnitude relation for Stephani model (Perlmutter
Sample A).}
\end{center}
\vspace{-0.5cm}
\label{fig:4}
\end{figure}

The lower line corresponds to standard Einstein-de Sitter model,
the line in the middle to our spherically symmetric Stephani model, i.e.
${\cal M}=-3.37, \quad \Omega_{m 0}=0.40$ (this line is
inseparable from the $\Lambda$CDM (Perlmutter) model with $\Lambda=0.7$
and $\Omega_{m 0}=0.3$), the upper line to Stephani model with $\Omega_{m 0}=0.3$.
From Fig. 4 we can see that the Perlmutter model and our best fit
model are indistinguishable on the basis of the present avaliable data.

In  Table 1 we present results of analysis with ${\cal M}$ obtained as a best
fit for "classical" Perlmutter model with $\Omega_{m 0}=0.3,
\Omega_{\Lambda 0}=0.7$ (top line for each samples) and with marginalization
over ${\cal M}$ (second line for each samples).
The best fit has been obtained using Bayes technique. For ${\cal M}=-3.39$,
we obtain value of $\Omega_{m 0}=0.37$ ($\chi^2=96.25$).
Another good fit can be obtained with marginalization over ${\cal M}$. In that
case we obtain ${\cal M}=-3.37,\quad \Omega_{m 0}=0.40$.
We test our results for the sample of 54 supernovae (Perlmutter
sample C) and we realize that no significant differences were obtained. Note that in our model
the numerical value for $\Omega_{m 0}$ is larger than in the standard approach
with cosmological constant, where $\Omega_{m 0}\approx 0.3$ ($\Omega_{m 0}=0.29$
for sample A and $\Omega_{m 0}=0.28$ for sample C \cite{Perlmutter99}).

However, knowledge of the best-fit values alone has not sufficient scientific relevance,
if confidence levels for parameter intervals are not presented too.
Therefore, we carry out the model parameters estimation using the minimization
procedure, based on the likelihood method.
We could observe that results obtained by this two methods are almost identical.
At the confidence level of $68.3~\%$ we obtain limits for values of parameters
${\cal M}$ and $\Omega_{m 0}$ separately for samples A and C (Table 2).
For example for Perlmutter sample A (with marginalization over ${\cal M}$) we
obtain that ${\cal M}=-3.37$ and $\Omega_{m 0}=0.39^{+0.10}_{-0.08}$
while for ${\cal M}=-3.39$ we have $\Omega_{m 0}=0.37^{+0.05}_{-0.05}$

Varying $\Omega_{m 0}$ and $\cal M$ we can find best-fit confidence regions in
the ($\Omega_{m 0}, {\cal M}$) plane for our Perlmutter supernovae sample.
In Fig. 5 we plot two confidence regions corresponding to levels of
$95.5$ (outer line) and of $68.3$ (inner line) respectively. Since $\cal M$ is
related to the absolute luminosity $M$ and the Hubble constant $H_0$ (see
relation (26)), knowing $M$ we can estimate $H_0$. For example
assuming that $0.25 < \Omega_{m 0} < 0.40$ (what is generally accepted) and
$M=-19.32$ we find that $61.9~{\rm km/sMpc}<H_0 <69.2~{\rm km/sMpc}$.

\begin{figure}[h]
\begin{center}
\includegraphics[width=0.6\textwidth]{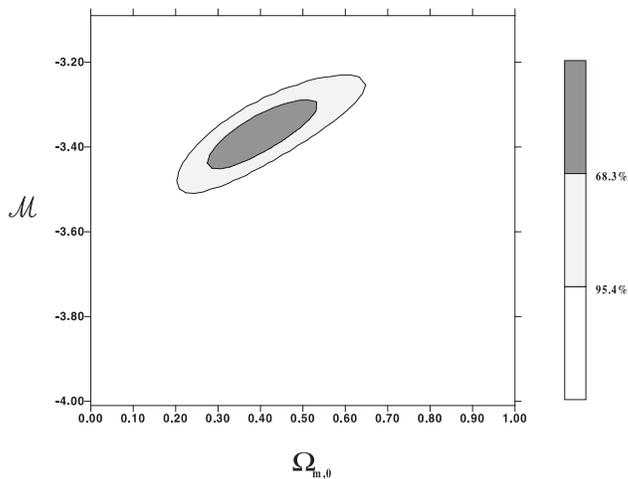}
\vspace{-0.5cm}
\caption{Confidence levels on the plane $(\Omega_{m 0}, {\cal M})$.
(Perlmutter Sample A). The figure shows of the preferred value of
$\Omega_{m 0}$ and ${\cal M}$ at the confidence of level of 68.3~\%
and 95.4~\%.}
\end{center}
\vspace{-0.5cm}
\label{fig:5}
\end{figure}

In Fig. 6 we show the levels of constant $\chi^{2}$ on the plane
$(\Omega_{m 0}, {\cal M})$. In this procedure we find the minimal value of
$\chi^2$, i.e., we consider best-fit values.
The figure shows the preferred values of $\Omega_{m 0}$ and ${\cal M}$.

\begin{figure}[h]
\begin{center}
\includegraphics[width=0.6\textwidth]{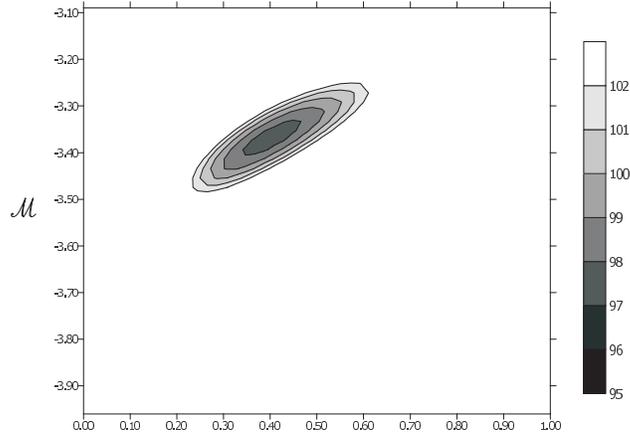}
\vspace{-0.5cm}
\caption{Levels of constant $\chi^{2}$ on the plane
$(\Omega_{m 0}, {\cal M})$
The figure shows the preferred value of
$(\Omega_{m 0}, {\cal M})$.}
\end{center}
\label{fig:6}
\vspace{-0.5cm}
\end{figure}

For a deeper statistical analysis of the Stephani model
in explaining the currently accelerating universe we consider 1D plot
of the density distribution of $\Omega_{{\rm non}0}$. From this
analysis one can obtain the limits at the $1\sigma$ or $2\sigma$  level.
Fig. 7 shows the density distribution for $\Omega_{{\rm non}0}$
in the Stephani model. This distribution is
obtained from the marginalization over $\mathcal{M}$.
We obtain as a best fit value $\Omega_{{\rm non}0} = 0.61$.
One can conclude that at the confidence level of $68.3~\%$
$\Omega_{{\rm non}0} \ge 0.51$ and $\Omega_{{\rm non}0} \le 0.70$ while
at the confidence level of $95.4~\%$ we obtain that $\Omega_{{\rm non}0}
\epsilon (0.41, 0.77)$. Fig. 8 shows the 1 dimensional density
distribution for $\mathcal{M}$.

\begin{figure}[h]
\begin{center}
\includegraphics[width=0.6\textwidth]{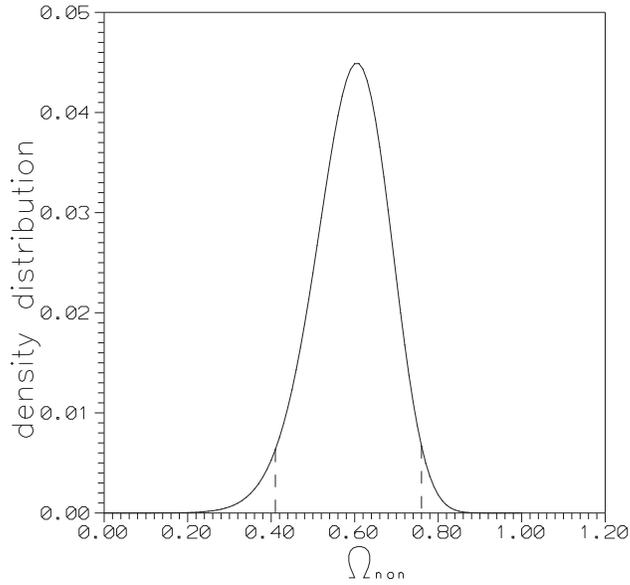}
\vspace{-0.1cm}
\caption{The density distribution for $\Omega_{{\rm non}0}$ in the Stephani
model (Perlmutter Sample A). $\Omega_{{\rm non} 0} > 0.41$,
$\Omega_{{\rm non}0} < 0.77$ at the confidence level of $95.4~\%$.}
\end{center}
\vspace{-0.8cm}
\label{fig:7}
\end{figure}
\clearpage

\begin{figure}[h]
\begin{center}
\includegraphics[width=0.6\textwidth]{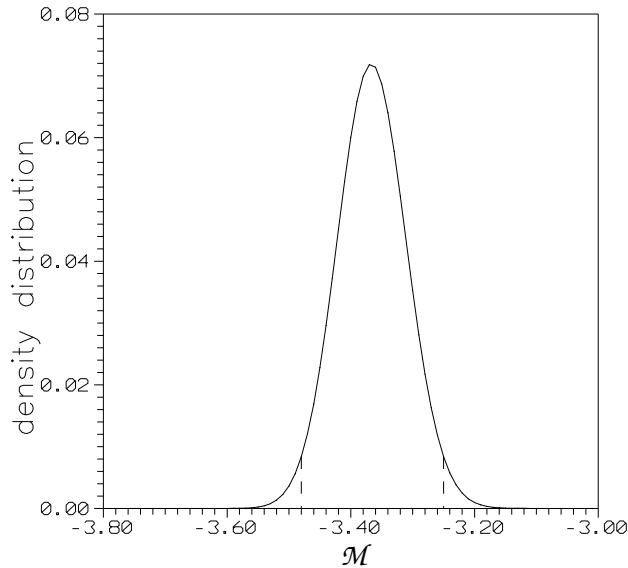}
\vspace{-0.1cm}
\caption{The density distribution for ${\cal M}$ in the Stephani
model (Perlmutter Sample A). The confidence level of $95.4~\%$ is
marked in the figure (${\cal M} \epsilon (-3.48,-3.25)$).}
\end{center}
\vspace{-0.8cm}
\label{fig:8}
\end{figure}

\section{Statistical analysis with the Knop and Tonry samples.}

Because the Perlmutter sample was completed four years ago, it would be
interesting to use newer supernovae observations. Lately Knop {\it et al.} \cite{Knop03}
have reegzamined the Permutter sample with host-galaxy extinction correctly
applied. They chose from the Perlmutter sample these supernovae which were
the more securely spectrally identified as type Ia and have reasonable colour
measurements. They also included eleven new high redshift supernovae and
a well known sample with low redshift supernowae.

We have also decided to test our  model using this new sample of supernovae.
The mentioned authors distinguished  few subsets of supernovae from this sample.
We consider two of them. The first is a subset of 58 supernovae with extinction
correction (Knop subsample 6; hereafter K6) and the second one a sample of 54 supernovae
with low extinction (Knop subsample 3; hereafter K3). Sample C and K3 are
similarly constructed because both contain only low extinction supernovae.

Another sample was presented by Tonry {\it et al.} \cite{Tonry03} who
collected a large number of supernovae published by different authors
and added eight new high redshift SN Ia. This sample of 230 Sne Ia
was recalibrated with consistent zero point. Whenever it was possible
the extinctions estimates and distance fitting were recomputed. However, none
of the methods was able to apply to all supernovae (for details see
Table~8 in \cite{Tonry03}). This sample was improved by Barris
who added 23 high redshift supernovae including 15 at $z \ge 0.7$ doubling the
published number of object at this redshifts \cite{Barris03}.

Despite of the mentioned above problems, the analysis of our model using
this sample of supernovae could be interesting. We decide to analyse
four subsamples. First, the full Tonry/Barris sample of 253 SNe Ia
(hereafter sample TBa) is considered. The sample of 218 SNe Ia (hereafter
sample TBb) consists of low extinction supernovae only (median $V$ band
extinction $A_V<0.5$).
Because the Tonry sample has a lot of outliers especially in low redshift,
we separately analysed the sample where all low redshift ($z<0.01$) supernovae
are excluded. This sample again contains 218 SN Ia, but they are different
than that belonging to the sample TBb (hereafter sample TBc).
In the sample of 193 SN Ia all supernovae with low redshift and and high
extinction are omitted (hereafter sample TBd).

\begin{figure}[h]
\begin{center}
\includegraphics[width=0.6\textwidth]{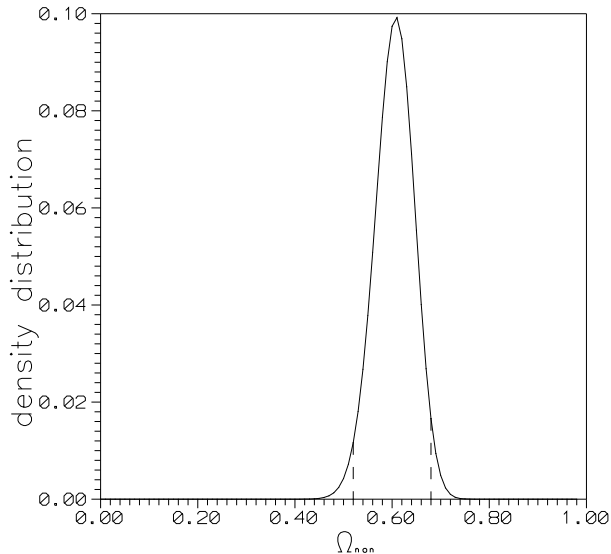}
\vspace{-0.5cm}
\caption{The density distribution for $\Omega_{{\rm non} 0}$ in the Stephani
model (Tonry/Barris sample TBa). $\Omega_{{\rm non} 0} \epsilon (0.52, 0.68)$
at the confidence level of $95.4~\%$.}
\end{center}
\label{fig:9}
\vspace{-0.5cm}
\end{figure}

Tonry and Barris \cite {Tonry03,Barris03} presented redshift and luminosity
distance observations for their sample of supernovae. Therefore,
Eqs. (25) and (26) should be modified \cite{Williams03}:
\begin{equation}
m-M = 5\log_{10}(\mathcal{D}_L)_{\mathrm{Tonry}}-5\log_{10}65 + 25
\end{equation}
and
\begin{equation}
\mathcal{M}=-5\log_{10}H_0+25.
\end{equation}
For $H_0=65$ km s$^{-1}$ Mpc$^{-1}$ we obtain $\mathcal{M}=15.935$.

The results obtained with the new sample is very similar to that obtained
with Perlmutter sample but now errors decrease. The results with Tonry
sample are almost identical to that with Perlmutter sample.
For the sample TBa we obtain that $\Omega_{m 0}=0.39$, while for $\lambda$CDM
model $\Omega_{m 0}=0.32$ (for the sample TBd $\Omega_{m 0}=0.33$
\cite{Barris03}).
Using Tonry sample we obtain at the confidence level of $95~\%$  limit for
$\Omega_{\rm non} \epsilon (0.52,0.68)$ (see Fig.~9).
However with the Knop sample we obtain the value $\Omega_{m 0}$ more closed
to value 0.3, what was obtained from CMBR and extragalactic
data \cite{Peebles03,Lahav02}, than in the previous case. However,
it should be noted, that for $\Lambda$CDM model we obtain for
the sample K3 $\Omega_{m 0}=0.25$ while for the sample K6 $\Omega_{m 0}=0.28$.
It means, that our conclusion that the numerical value for $\Omega_{m 0}$
in our model is larger than in the standard approach ($\Lambda$CDM model)
is still valid.

\begin{figure}[h]
\begin{center}
\includegraphics[width=0.75\textwidth]{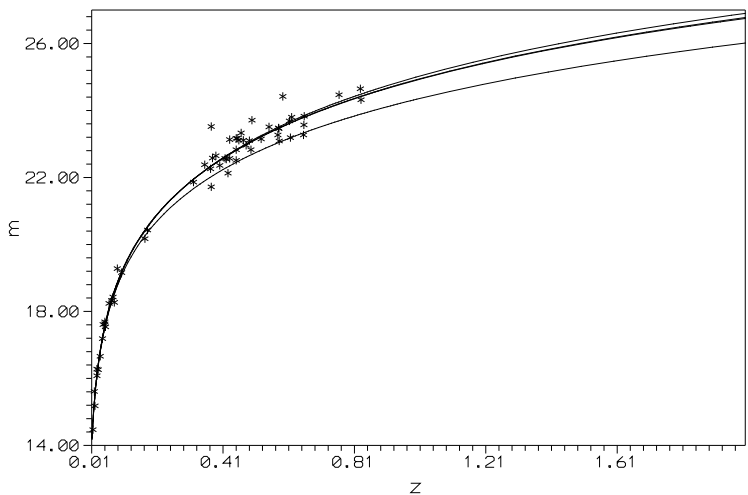}
\vspace{-0.5cm}
\caption{The Redshift-magnitude relation for the Stephani model (Knop
Sample K3).}
\end{center}
\label{fig:10}
\vspace{-0.5cm}
\end{figure}

In Fig.~10 we present plots of redshift-magnitude relations
for Stephani model Perlmutter sample A. The lower line corresponds to
the standard Einstein-de Sitter model,
the line in the middle to our spherically symmetric Stephani model, i.e.
${\cal M}=-3.46, \quad \Omega_{m 0}=0.32$ (this line is
inseparable from the $\Lambda$CDM model with $\Lambda=0.75$ and $\Omega_{m 0}=0.25$),
 the upper line to Stephani model with $\Omega_{m 0}=0.25$.
From Fig.~10 we can see that the $\Lambda$CDM model and our best fit
model are indistinguishable also on the basis of the Knop sample.

\begin{figure}[h]
\begin{center}
\includegraphics[width=0.6\textwidth]{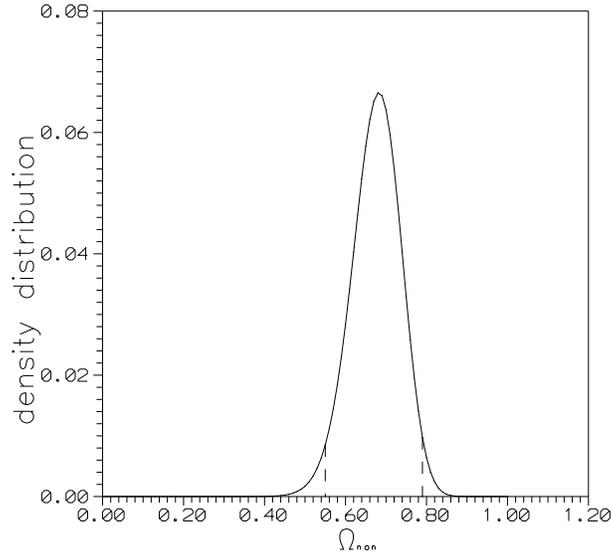}
\vspace{-0.5cm}
\caption{The density distribution for $\Omega_{{\rm non} 0}$ in the Stephani
model (Knop sample K3). $\Omega_{{\rm non}0} \epsilon (0.55, 0.79)$
at the confidence level of $95.4~\%$.}
\end{center}
\label{fig:11}
\vspace{-0.5cm}
\end{figure}

\begin{figure}[h]
\begin{center}
\includegraphics[width=0.6\textwidth]{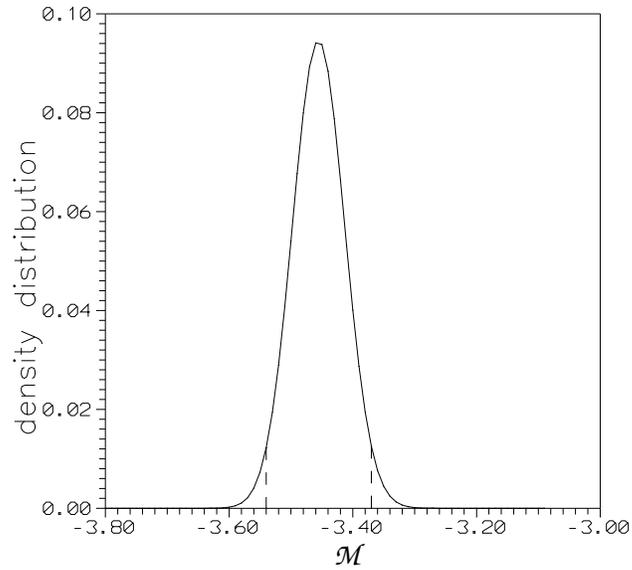}
\vspace{-0.4cm}
\caption{The density distribution for ${\cal M}$ in the Stephani
model (Knop Sample K3). ${\cal M} \epsilon (-3.54,-3.37)$) at the confidence
level of $95.4~\%$.}
\end{center}
\label{fig:12}
\vspace{-0.5cm}
\end{figure}
\clearpage

Using the Knop sample (Figs~11  and 12) we obtain  $\Omega_{\rm non}
\epsilon (0.55,0.79)$ at the confidence level of $95~\%$.
Because Knop discuss very carefully extinction correction
and as a result his sample has extinction  correctly applied,
we think that using the limit obtained from the Knop's sample is the most
appropriate. Our results show that
Stephani  model is consistent with SNIa data at the $95~\%$ confidence level.
Our study shows that Stephani model is very good fit to latest supernovae data
and should be treated as a alternative to $\Lambda$CDM (Perlmutter) model.

\section{CMB peaks in the Stephani model}

The CMB peaks arise from acoustic oscillations of the primeval plasma.
Physically these oscillations represent hot and cold spots. Thus, the
wavelength of the perturbation which contributes the most to the density
distribution at the time of the last scattering corresponds to
a peak in the power spectrum. In the Legendre multipole space this
corresponds to the angle subtended by the sound horizon at the last
scattering epoch. Higher harmonics of the principal oscillations, which
oscillated more than once, correspond to secondary peaks.

It is well known that the locations of the peaks are very sensitive to
the variations in the parameters of the model. Therefore, it can be
used as a sensitive probe to constraint the cosmological parameters and
discriminate among various models.

The locations of the peaks are set by the acoustic scale $l_{A}$ which can
be defined in terms of an angle $\theta_{A}$ subtended by the sound horizon
at the last scattering surface. In a general case calculation of the angle
$\theta$ corresponding to acoustic or particle horizon existing at
the recombination epoch is not an easy task even in the Friedmann model. One of
the way to find $\theta$ is to solve an appropriate Euler-Lagrange problem
(\cite{narl83}, \cite{wein72}).
If the angle $\theta_{A}$ is small (in general it is small, because $c_s <<
c$), hence the acoustic scale $l_{A} = \pi/\theta_{A}$ is given by
\begin{equation}
l_{A} = \pi \frac{\int^{1}_{x_{\mathrm{dec}}} dx/h(x)}
{\int^{x_{\mathrm{dec}}}_{0} c_{s}(x)dx/h(x)},
\end{equation}
where
\begin{equation}
h(x) = \sqrt{\Omega_{r 0}
+ \Omega_{m 0} x+\Omega_{{\rm non}0} x^3}
\end{equation}
and the relation between $z$ and $x$ is given by the equation
(31). $c_{s}(x)$ is a speed of sound in the plasma and varies with the expansion
(we assume additionally presence of radiation in the model and $\Omega_{r 0}$
is an energy density parameter corresponding to radiation at the present epoch).
Similarly $\Omega_{m 0}$ corresponds to nonrelativistic matter
and $\Omega_{{\rm non}0}$ to nonhomogeneity. The sound
velocity can be calculated from the formula
\begin{equation}
c_{s}^{2}(x) =
\frac{\frac{4}{3}\Omega_{r 0} - \frac{2}{3} \Omega_{{\rm non}0}x^3}
{4\Omega_{r 0}+ 3\Omega_{m 0}x+\Omega_{{\rm non}0}x^3}
\end{equation}
In the case without cosmological term the parameters
$\Omega_{m 0}$ and $\Omega_{{\rm non}0}$ are not
independent, and $\Omega_{{\rm non}0}$ can be expressed as
\begin{equation}
\Omega_{{\rm non}0}= 1-\Omega_{r 0} -
\Omega_{m 0}.
\end{equation}
In the model of primeval plasma, there is a simple relation
\begin{equation}
l_{m} \approx l_{A}(m - \phi_{m})
\end{equation}
between the location of the $m$-th peak and the acoustic scale
\cite{Doran01,Hu01}. The prior assumptions in our calculations are
 $\Omega_{r 0} = 9.89 \cdot 10^{-5}$,
$\Omega_{b 0} = 0.05$ (baryonic matter energy density),
and the spectral index for initial
density perturbations is $n = 1$. Moreover we put for the present value of
the Hubble parameter $H_0 = 65$ km/sMpc.

The phase shift is caused by the pre-recombination physics (plasma driving
effect) and hence, is not significantly influenced by the Stephani term
at that epoch. Therefore the phase shift $\phi_{m}$ can be taken from
standard cosmology \cite{Hu01}
\begin{equation}
\phi_{m} \approx 0.267 \left[ \frac{r(z_{\mathrm{dec}})}{0.3} \right]^{0.1},
\end{equation}
where $\Omega_{b 0} h^{2} = 0.02$, $r(z_{\mathrm{dec}}) \equiv
\rho_{r}(z_{\mathrm{dec}})/\rho_{m}(z_{\mathrm{dec}})
= \Omega_{r 0}/\Omega_{m 0}x_{\mathrm{dec}}$.
Radiation energy is composed of two components: electromagnetic
energy and neutrino energy $\Omega_{r 0} = \Omega_{\gamma 0} +
\Omega_{\nu 0}$, $\Omega_{\gamma 0} = 2.48h^{-2} \cdot 10^{-5}$,
$\Omega_{\nu 0} = 1.7h^{-2} \cdot 10^{-5}$. $r(z_{\mathrm{dec}})$
is the ratio of radiation to matter densities at the surface of last
scattering.

In the Friedman models ($\Omega_{{\rm non}0}=0$) with possible
presence of cosmological constant we obtain:

\begin{tabular}{lll}
$\Omega_{m 0} = 1.0$
& $\Omega_{{\rm non}0} = 0. \colon$ & $l_{\mathrm{peak},1} = 203$,
$l_{\mathrm{peak},2} = 471$,  $l_{\mathrm{peak},3} = 739$, \\
$\Omega_{m 0} = 0.3$
& $\Omega_{\Lambda} = 0.7 \colon$ & $l_{\mathrm{peak},1} = 225$,
$l_{\mathrm{peak},2} = 536$,  $l_{\mathrm{peak},3} = 847$.
\end{tabular}

The influence of $\Omega_{{\rm non}0}$ on the location of the peaks results in
shifting them towards higher values of $l$ in comparison to the model with
$\Omega_{r 0}$. For example, for $\Omega_{b 0} = 0.05$,
$h = 0.65$, the different choices of $\Omega_{{\rm non}0}$ yield

\begin{tabular}{lll}
$\Omega_{m 0} = 0.4$
& $\Omega_{{\rm non}0} = 0.6 \colon$ & $l_{\mathrm{peak},1} = 212$,
$l_{\mathrm{peak},2} = 500$,  $l_{\mathrm{peak},3} = 789$, \\
$\Omega_{m 0} = 0.3$
& $\Omega_{{\rm non}0} = 0.7 \colon$ & $l_{\mathrm{peak},1} = 216$,
$l_{\mathrm{peak},2} = 514$,  $l_{\mathrm{peak},3} = 812$.
\end{tabular}

On the other hand from the Boomerang observations \cite{deBernardis02}
we obtain $l_{\mathrm{peak},1} = 200\div 223$, $l_{\mathrm{peak},2} = 509\div 561$.
We also compare the results from our model to recent bounds on the
location of the first two peaks obtained by WMAP experiment
\cite{Spergel03,Page03}. Namely $l_{\mathrm{peak},1} = 220.1 \pm 0.8$,
$l_{\mathrm{peak},2} = 546 \pm 10$, together with the bound of the location
of the third peak obtained by Boomerang experiment
$l_{\mathrm{peak},3} = 825^{+10}_{-13}$ which lead to quite strong constraints
on the model parameters. These constraints can be summarized as follows.
The Stephani  model is in agreement with observations and we conclude that
the influence of the term $\Omega_{{\rm non}0}$ is not very significant in
our case. However, phase shift $\phi$ is taken from the standard cosmology,
i.e., we assume that the contribution from the "Stephani nonhomogeneity"
is insignificant at the pre-recombination epoch. If this assumption is not
valid then the limit from CMB will change.

It should be also noted that the Big-Bang Nucleosynthesis (BBN) is a very
well tested area of cosmology and does not allow for significant deviation
from the standard expansion law, apart from very early times before the onset
of the BBN. The consistency with the BBN seems to be necessary in the Stephani model.
We consider only non-relativistic matter $\rho$ in the Stephani model.
In this case we could approximate that $\Omega_{{\rm non}0}$ term scales like
$(1+z)$. It is clear that the contribution of the "Stephani nonhomogeneity term"
cannot dominate over the standard radiation  term before the onset of BBN,
i.e., for $z \cong 10^8$ and as a result has no effect on the BBN.

\section{Conclusions}

Our investigations show that the Stephani model is an excellent fit to both Perlmutter
data points and currently available Knop's  data points.
Summarizing, relatively large value of $\Omega_{{\rm non}0}=0.61^{+0.08}_{-0.10}$
can explain supernovae data of Perlmutter. In the range of observed redshifts
of supernovae ($z<1$) the curve corresponding to $m-z$ relation in our model
is almost indistinguishable from the $\Lambda$CDM (Perlmutter) model with
cosmological constant. With future data from SNAP the error bars in the
extimation of the model parameters will be reduced significantly.
Together with new limits for $\Omega_{m,0}$ (obtained from extragalactic data)
it will be possible discriminate between Stephani model
and $\Lambda$CDM (Perlmutter) model.

Locally our spherically symmetric Stephani universe is indistinguishable
from the Friedman one (the same equation of state, the same local geometry),
hence local physical cosmology in both models must be the same. In other words
standard scenario of the evolution of the universe remains valid. Especially
in the early universe effects of non-homogeneity become negligible. This is
easily seen if we evaluate the energy density parameter
$\Omega_\alpha$ for matter as a function of the scale factor $R$
\begin{equation}
\Omega_\alpha(R)={1\over 1-3\beta C_\alpha R^{2+3\alpha}}.
\end{equation}
In the early universe relativistic matter dominates ($\alpha=1/3$) and from
the above relation it follows that $\Omega_\alpha(R)$ tends very quickly to $1$ for
$R \to 0$. It means that $\Omega_{\rm non}(R) \to 0$. This justifies
application of standard physical cosmology (including e.g. primordial
nucleosynthesis) for our nonhomogeneous model.

Freese and Lewis in \cite{Freese02} have recently proposed interesting
alternative model explaining the currently accelerating Universe.
In this model which is called the Cardassian model the standard
Friedmann-Robertson-Walker (FRW) equation is modified by the presence
of an additional term $\rho^n$, namely
\begin{equation}
H^2 = \frac{\rho}{3} + B\rho^n,
\end{equation}
where
$\rho$ is a total energy density (matter $+$ radiation), and $B$ is a positive
constant. This proposal seems to be attractive because the expansion of the
universe is accelerated automatically due to the presence of the additional
term (if we put $B=0$ then the standard FRW equation is recovered) without
postulating existence of unknown form of dark energy.

Note that putting $\rho\equiv 3\rho_{m 0}/R^3$ (and hence $\alpha\equiv 0$)
and $n=1/3$ and $B\equiv -3\beta/\sqrt{3\rho_{m 0}}$ the above equation could
be rewritten in the form:
\begin{equation}
H^2 = \frac{\rho}{3} + B\rho^n={\rho_{m 0} \over R^3}+{(-3\beta)\over R}.
\end{equation}
Comparing to (1) we realize that we obtain correspondence between Cardassian
and Stephani models.

When for simplicity we assume that the energy density parameter for
radiation matter vanishes ($\Omega_{r 0}=0$) we recover the model
analysed in our previous paper \cite{Godlowski04}. We note that  considered
by us spherically symmetric Stephani model is a special realization or in other
words is dynamically equivalent to the Cardassian model for n=1/3. It is
interesting that this value of $n$ parameter is reasonable from
the observational joint test analysis of both CMB from and  SNIa \cite{Sen03}.

Let us note that in our model there is the cosmic coincidence problem, i.e.
problem why did the nonhomogeneity start to dominate the present evolution of
the Universe only fairly recently? There is no satisfactory solution of this
problem and some  other face of fine tuning coincidence is required.

\clearpage

\begin{table}[h]
\noindent
\caption{Results of the statistical analysis of the model
using the best fit with minimum $\chi^2$ (denoted by BF).
The case in which we marginalize over $\mathcal{M}$ is denoted
by $\mathcal{M}$.}
\vspace{0.2cm}
\label{tab:1}
\begin{center}
\begin{tabular}{@{}p{1.5cm}rrrrrr}
\hline \hline
sample &  $\Omega_{m 0}$ & $\Omega_{\mathrm{non} 0}$ &
$\mathcal{M}$ & $\chi^2$ & method \\
\hline
  A   &  0.37 & 0.63 & -3.39 & 96.3 & BF               \\
      &  0.40 & 0.60 & -3.37 & 96.1 & $\mathcal{M}$, BF\\
  C   &  0.36 & 0.64 & -3.42 & 53.4 & BF               \\
      &  0.36 & 0.64 & -3.42 & 53.4 & $\mathcal{M}$, BF\\
  K3  &  0.29 & 0.71 & -3.48 & 61.3 & BF               \\
      &  0.32 & 0.68 & -3.46 & 61.0 & $\mathcal{M}$, BF\\
  K6  &  0.36 & 0.64 & -3.53 & 56.3 & BF               \\
      &  0.36 & 0.64 & -3.51 & 56.0 & $\mathcal{M}$, BF\\
  TBa &  0.38 & 0.62 & 15.905&262.3 & BF               \\
      &  0.39 & 0.61 & 15.915&262.2 & $\mathcal{M}$, BF\\
  TBb &  0.39 & 0.61 & 15.925&204.8 & BF               \\
      &  0.39 & 0.61 & 15.925&204.8 & $\mathcal{M}$, BF\\
  TBc &  0.39 & 0.61 & 15.915&229.3 & BF               \\
      &  0.41 & 0.59 & 15.925&229.2 & $\mathcal{M}$, BF\\
  TBd &  0.39 & 0.61 & 15.925&192.7 & BF               \\
      &  0.39 & 0.61 & 15.925&192.7 & $\mathcal{M}$, BF\\
\hline
\end{tabular}
\end{center}
\end{table}

\begin{table}[h]
\caption{Results of the statistical analysis of the model
using the likelihood method (denoted by L).
The case in which we marginalize over $\mathcal{M}$ is denoted
by $\mathcal{M}$.}
\label{tab:2}
\vspace{0.2cm}
\begin{center}
\begin{tabular}{@{}p{1.5cm}rrrr}
\hline \hline
sample & $\Omega_{m 0}$ & $\mathcal{M}$ & method \\
\hline
   A  &  $0.37^{+0.05}_{-0.05}$ &$ -3.39              $ & L             \\
   &  $0.39^{+0.10}_{-0.08}$ &$ -3.39^{+0.06}_{-0.05} $ & L$\mathcal{M}$\\
C  &  $0.36^{+0.05}_{-0.05}$ &$ -3.42                 $ & L             \\
   &  $0.36^{+0.10}_{-0.08}$ &$ -3.42^{+0.05}_{-0.05} $ & L$\mathcal{M}$\\
K3 &  $0.29^{+0.04}_{-0.03}$ &$ -3.48                 $ & L             \\
   &  $0.29^{+0.06}_{-0.06}$ &$ -3.46^{+0.05}_{-0.04} $ & L$\mathcal{M}$\\
K6 &  $0.31^{+0.06}_{-0.06}$ &$ -3.53                 $ & L             \\
   &  $0.33^{+0.08}_{-0.07}$ &$ -3.51^{+0.03}_{-0.03} $ & L$\mathcal{M}$\\
TBa&  $0.38^{+0.03}_{-0.03}$ &$ 15.905                $ & L             \\
   &  $0.39^{+0.04}_{-0.04}$ &$ 15.915^{+0.02}_{-0.02}$ & L$\mathcal{M}$\\
TBb&  $0.39^{+0.03}_{-0.03}$ &$ 15.925                $ & L             \\
   &  $0.39^{+0.04}_{-0.04}$ &$ 15.925^{+0.02}_{-0.02}$ & L$\mathcal{M}$\\
TBc&  $0.39^{+0.02}_{-0.02}$ &$ 15.915                $ & L             \\
   &  $0.40^{+0.04}_{-0.04}$ &$ 15.925^{+0.02}_{-0.02}$ & L$\mathcal{M}$\\
TBd&  $0.39^{+0.03}_{-0.03}$ &$ 15.925                $ & L             \\
   &  $0.39^{+0.04}_{-0.04}$ &$ 15.935^{+0.02}_{-0.02}$ & L$\mathcal{M}$\\
\hline
\end{tabular}
\end{center}
\end{table}

\end{document}